# A new prediction method of unsteady wake flow by the hybrid deep neural network

Renkun Han [1, 2], Yixing Wang [1, 2], Yang Zhang [1] and Gang Chen [1, 2]

1 State Key Laboratory for Strength and Vibration of Mechanical Structures, School of Aerospace Engineering, Xi'an Jiaotong University, Xi'an 710049, China

2 Shannxi Key Laboratory for Environment and Control of Flight Vehicle, Xi'an Jiaotong University, Xi'an 710049, China

**Abstract:** The fast and accurate prediction of unsteady flow becomes a serious challenge in fluid dynamics, due to the high-dimensional and nonlinear characteristic. A novel hybrid deep neural network (DNN) architecture was designed to capture the unsteady flow spatio-temporal features directly from the high-dimensional unsteady flow fields. The hybrid deep neural network is constituted by the convolutional neural network (CNN), convolutional Long Short Term Memory neural network (ConvLSTM) and deconvolutional neural network (DeCNN). The flow around a cylinder at various Reynolds numbers and the flow around an airfoil at higher Reynolds number are carried out to establish the datasets used to train the networks separately. The trained hybrid DNNs were then tested by the prediction of the flow fields at future occasions. The predicted flow fields using the trained hybrid DNNs are in good agreement with the flow fields calculated directly by the computational fluid dynamic solver.

## 1. Introduction

High-fidelity modeling of unsteady flows is one of the major challenges in computational physics, with diverse applications in engineering. The rising popularity of high fidelity computational fluid dynamics (CFD) techniques have made significant inroads into this problem. However, such direct numerical simulations with billions of degrees of freedom in order to resolve the energetically relevant spatial and temporal scales lead to the time-consuming computations. Therefore, the data-driven low-dimensional models that can capture the main dynamic characteristics of the unsteady dynamic systems with good efficiency and accruacy were proposed [1].

Reduced order modeling (ROM), as one of the model identification methods, has proven to be a powerful tool to reduce the complexity and large dimensional size of the full discretised dynamical system. ROM, such as the proper orthogonal decomposition (POD) [2] and dynamic mode decomposition (DMD) [3], offers the potential to simulate physical and dynamic systems with substantially increased computational efficiency while maintaining reasonable accuracy. Lots of researches have been carried out to analyze flow fields with low-dimensional representations using these methods. Since most of these investigations are linear or weakly nonlinear methods with some strong assumptions, the type of flow fields that they can analyze properly is limited.

Deep learning technology is a recent advancement in artificial neural networks which is capable of finding more complex and hidden information from the big data. It has the advantage of learning the non-linear system with multiple levels of representation data [4]. Recently, it has resulted in breakthroughs in various areas like image processing [5], speech recognition [6] and disease diagnosis [7]. More recently, there has been a surge of interest in deep learning applications to fluid mechanics. A few attempts to apply deep learning method to fluid dynamics for turbulence modeling with Reynolds-averaged Navier-Stokes (RANS) simulations have been conducted, such as J. Ling et al [8], Wu J et al [9] and R. Maulik et al [10]. These method could reduce computational costs of RANS and increase accuracy by utilizing neural networks to learn Reynolds stress closures.

However, there exist applications where solely parameterizations of turbulence may not be ideal approaches. Especially for unsteady flows involving fluid-structure interactions, it pose serious challenges of fast and accurate prediction due to the richness and complexity of nonlinear coupled physics. Even a simple configuration of a coupled fluid-structure system can exhibit complex spatial-temporal dynamics and synchronization as functions of physical parameters and geometric variations. Therefore, there is an urgent need to model the spatio-temporal dynamics of unsteady flows, which are physically consistent, at a low computational cost. Compared with the traditional ROM, the ROM using deep learning method can capture more nonlinear features mapping data without loss of information while incorporating the appropriate degree of nonlinearity. The hybrid deep neural network is designed to achieve fast and accurate prediction of unsteady flow fields, which is very important for flow control.

The deep learning method is good at finding more complex and hidden information from nonlinear system, so deep learning is well suited for constructing a reduced order model of unsteady flow. In the work of Wang [11] and Noriyasu Omata et al [12], deep convolutional autoencoder data-driven nonlinear low-dimensional representation method was used for dimensionality reduction of unsteady flow fields. Kai Fukami et al [13] used the CNN and the hybrid Downsampled Skip-Connection Multi-Scale models to perform super-resolution analysis of grossly under-resolved low-dimensional turbulent flow field data to reconstruct the high-resolution high-dimensional flow field. Jin et al [14] predicted the velocity field around a cylinder by fusion convolutional neural networks using measurements of the pressure field on the cylinder. In these works, the deep learning method is only used to do the dimensionality reduction or reconstruct the current time flow field, which is not able to model the spatio-temporal flow dynamics and predict the unsteady advancing flow fields at future occasions.

To introduce time effects into neural networks for fluid dynamics, A T. Mohan et al [15] constructed a ROM by combining the POD method and the LSTM network, which was used to predict the time coefficients of the flow POD modes. They [16] also used the combination of CNN and the ConvLSTM for dimensionality reduction and spatio-temporal modeling of the three-dimensional dynamics of turbulence. Sangseung Lee et al [17, 18] used the generative adversarial networks (GAN) to predict unsteady flow fields at future occasions. In this paper we will take the advantage of the latest research progress in the reverse of CNN, LSTM and DeCNN with the goal of a) Dimensionality reduction of high dimensional unsteady flow datasets, and b) Learning the attractor of the governing dynamical system to model its spatio-temporal dynamics at a low computational cost.

Different from previous work, we try to use one neural network to reduce the dimensionality of the unsteady flow fields and capture the spatio-temporal flow dynamics characteristics simultaneously for the first time. More specifically, a novel hybrid deep neural network is designed to capture the complex mapping directly from the high-dimensional flow fields without using any explicit intermediate dimensionality reduction method and predict the flow fields at future occasions based on captured features of flow fields at previous occasions. The structure of this article is as follows: Sec. 2 introduces the architecture of the proposed hybrid deep neural network architecture. Then, in Sec. 3, the method for constructing flow field datasets and the neural network training algorithm are explained. Sec 4 evaluates the general performance of the proposed hybrid deep neural network. Finally, a summary is provided in Sec. 5.

## 2. The architecture of the hybrid deep neural network

### 2.1. Architectural design of the hybrid deep neural network

As explained in the previous section, the goal of this work is to reduce the dimension of high dimensional unsteady flow data and lean its spatio-temporal dynamic characteristics with deep learning neural network Therefore, a hybrid deep neural network architecture composed of CNN layers, ConvLSTM layer and DeCNN layers is designed to capture accurate spatial-temporal features of unsteady flows, as is shown in Fig. 1.

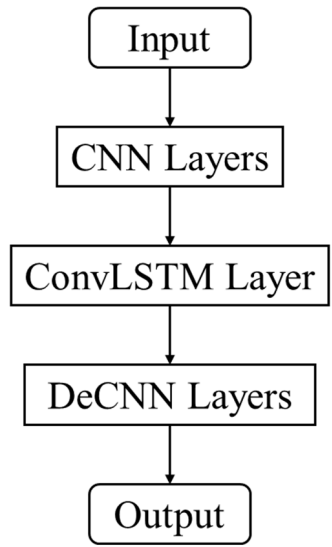

Figure 1. The architecture of the hybrid deep neural network

The CNN layers, consisting of six convolutional layers, are designed to capture the complex features directly from the high-dimensional input fields, and represent it in low-dimensional form. Several features of each time step flow field are obtained by CNN layers. Before the ConvLSTM layer, a reshape layer is settled to reshape the CNN layers' output matrix to the shape ConvLSTM layer's input should be, turning the batch size of CNN layers to the time steps of the ConvLSTM layer. Then, a ConvLSTM layer is designed to capture temporal features between low-dimensional features representation and predict the feature map of flow fields at future occasions.

For each kind flow field feature map, the ConvLSTM layer predict the flow field feature map at future occasions based on corresponding feature maps of flow fields at previous occasions. The DeCNN layers copies the architecture of the CNN layers and reverses it. The DeCNN layers are used to represent the predicted low-dimensional features to high-dimensional output field, with the same dimension as input field. All the layers and connections used to build the hybrid deep neural network are described in detail as follows.

## 2.2 Convolutional layer

The convolutional layer's parameters consist of a set of learnable filters (or kernels), which have a small receptive field, but extend through the full depth of the input volume. During the forward pass, each filter is convolved across the width and height of the input volume, computing the dot product between the entries of the filter and the input and producing a 2-dimensional activation map of that filter. As a result, the network learns filters that activate when it detects some specific type of feature at some spatial position in the input. Stacking the activation maps for all filters along the depth dimension forms the full output volume of the convolution layer.

The convolutional layer and the nonlinear activation function compute the output feature map of the previous layer via a convolutional, i.e., filtering, operator,

$$y_c = \sigma\left(k * x_c + b_c\right) \tag{1}$$

where $x_c$ is the input feature map of the convolutional layer, $*$ is the convolutional operator, $k$ is the learnable convolutional kernel, $b_c$ is the additive bias, $\sigma$ is the nonlinear activation function, and $y_c$ is the output feature map.

In this study, leaky rectified linear unit (PReLU) is used as the nonlinear activation function $\sigma$, i.e., since the negative value exists in the flow field. $a$ is a constant with a very small value, setted to 0.02 in this article

$$\sigma\left(x\right) = P\mathrm{Re}LU\left(x\right) = \max\left(ax, x\right) \tag{2}$$

## 2.3 Convolutional Long Short Term Memory layer

The LSTM networks are specialized at capturing the temporal characteristics. The typical LSTM cell contains three gates: the input gate, output gate and the forget gate. The LSTM regulates the flow of training information through these gates by selectively adding information (input gate), removing information (forget gate) or letting it through to the next cell (output gate). LSTMs utilize their internal memory, such that the predictions are conditional on the recent context in the input sequence, not what has just been presented as the current input to the network. For instance, to predict the realization at time $t_i$, LSTMs can learn from the data at $t_{i-1}$ and also at times $t_{i-k}$, where $k$ can be any number signifying the length of the prior sequence. In effect, $k$ represents the "memory" in the system i.e. the extent to which the outcome of the system depends on its previous realizations.

In traditional LSTMs, the input and hidden states consist of a one-dimensional vector, therefore a two dimensional input (such as an image or a data field) has to be resized to a single dimension. The "removal" of this dimensionality information fails to capture spatial correlations that may exist in such data, leading to increased prediction errors. In contrast, the ConvLSTM was proposed by X Shi [19] to process hidden and input states in two dimensions, thereby retaining spatial information in the data. ConvLSTM consists of a simple but powerful idea, that the gates have the same dimensionality of the input data. This enables us to provide a 2D image input and obtain 2D vectors cell state as outputs from the ConvLSTM cell.

The input gate is represented by $i$, output gate by $o$ and forget gate by $f$. The cell state is represented as $c$ and the cell output is given by $h$, while the cell input is denoted as $x$. Consider the equations of a ConvLSTM cell to compute its gates and states in Eqn 3.

$$
\begin{aligned}
f_t &= \sigma\left(W_{xf} * \chi_t + W_{hf} * h_{t-1} + W_{cf} \odot c_{t-1} + b_f\right) \\
i_t &= \sigma\left(W_{xi} * \chi_t + W_{hi} * h_{t-1} + W_{ci} \odot c_{t-1} + b_i\right) \\
c_t &= f_t \odot c_{t-1} + i_t \odot \tanh\left(W_{xc} * \chi_t + W_{hc} * h_{t-1} + b_c\right) \\
o_t &= \sigma\left(W_{xo} * \chi_t + W_{ho} * h_{t-1} + W_{co} \odot c_{t-1} + b_o\right) \\
h_t &= o_t \odot \tanh\left(c_t\right)
\end{aligned}
\tag{3}
$$

$W$ are the weights for each of the gates and * is the convolutional operator.

## 2.4 DeConvolutional layer

The concept of deconvolution is proposed by Zeiler [20, 21]. In his work, the deconvolution was used for the visualization of every network layer learning result. With the successful application of deconvolution in neural network visualization, it has been adopted by more and more work such as scene segmentation. In mathematics, deconvolution is an algorithm-based process used to reverse the effect of convolution on recorded data. The essence of deconvolution is convolution, but with an automatic zero padding is added before the convolution.

Deconvolutional layers multiply each element of the input with a filter (kernel) and sum over the resulting feature map, effectively swapping the forward and backward passes of a regular convolutional layer. The effect of using deconvolutional layers is to decode low-dimensional abstract features to a larger dimensional representation. It should be noted that the deconvolution cannot restore the matrix before convolution, and can only restore the size.

## 3. Training Method of the Hybrid Deep Neural Network

### 3.1 Dataset Constructions

Because CNN are developed from the field of computer vision, we consider the goal of predicting the flow fields at future occasions based on captured features of flow fields at previous occasions in a similar fashion to approaches considered in deep learning for image-to-image regression tasks. Therefore the dataset used for training and testing networks should be image like dataset. The flow field information value at each moment should be distributed over evenly distributed grid points, like pixels.

We use high-precision numerical simulation to calculate the dynamic process of unsteady flow field and record the flow field information at each moment. The original quantities in the incompressible Navier-Stokes equations are nondimensionalized as follows:

$$\mathrm{u}^* = \frac{u^i}{U_0}, \mathrm{v}^* = \frac{v^i}{U_0}, \mathrm{p}^* = \frac{p^i}{\rho U_0^2}, \mathrm{x}_i^* = \frac{x^i}{D}, \mathrm{y}_i^* = \frac{y^i}{D}, \mathrm{t}^* = \frac{t}{T} \qquad (4)$$

where $u^i$, $v^i$, $x^i$, $y^i$, $t$, $p^i$ are the dimensional velocity, length, time, and pressure, respectively, and $u^*$, $v^*$, $x^*$, $y^*$, $t^*$, $p^*$ are nondimensionalized velocity, length, time, and pressure by the incoming velocity $U_0$, the characteristic length $D$, the fluid density $\rho$, and vortex shedding frequency $1/T$.

A rectangular area should be chosen as the sampling area. Lattice like sampling points of $Nx \times Ny$ are placed in the space. Then, project the nondimensionalized flow field variables quantities onto the uniformly distributed grid. The values of the point inside the body are 0. Three-dimensional flow field variables ($p^*$, $u^*$ and $v^*$) are extracted at each sampling point. The $Nx \times Ny \times 3$ dimensional data that are extracted represent each instantaneous field. The obtained data is arranged in chronological order to obtain a dataset for training and testing the neural network.

## 3.2 Training Algorithm

The root mean square error (RMSE) is used to evaluate the model performance, i.e.

$$\text{RMSE}^t = \sqrt{\frac{\sum_{i=1}^{N}\left(\psi_i^t - \psi_{o,i}^t\right)^2}{N}} \qquad (5)$$

where $\psi_i^t$ and $\psi_{o,i}^t$ denote the predictions and numerical simulations at the node $i$ and the time level $t$, respectively, and $N$ represents the number of nodes on the full mesh.

Training the network is equivalent to minimizing the loss function in Eqn 5 to obtain the optimal all kernel parameters. The backpropagation method is implemented for training, which involves calculating the gradients of the loss function with respect to the learnable parameters. In the backpropagation method, the sensitivity of every layer is the variable to back propagate the loss in Eqn 5, which can be obtained in the literature [22]. Then, the gradients of the loss function with respect to the parameters of the network can be computed.

Adaptive moment estimation (Adam) is employed as the optimization algorithm for training the network [23]. In this algorithm, the exponential moving average is used to update the gradient vector and the squared gradient. Training of the network is carried out with the open-source software library TensorFlow [24].

## 4. Results and Discussions

## 4.1 DNNs training

We apply the hybrid deep neural network described in the previous sections on three representative experiments to illustrate the effectiveness of the approach to nonlinear model reduction and predicting the flow fields at future occasions based on flow fields at previous occasions. Three numerical simulations, namely flows past a cylinder at various Reynolds numbers and flow around an airfoil, were conducted. The first numerical experiment (Case 1) is the flow past a cylinder at Reynolds number Re

= 200; and the second experiment (Case 2) is the flow past a cylinder at Re = 4000; the last experiment (Case 3) is the flow past the NACA0012 airfoil at Re = 8000 and angle of attack α = 20°. In all experiments, the flow field data were computed and translated to deep learning datasets by the same way.

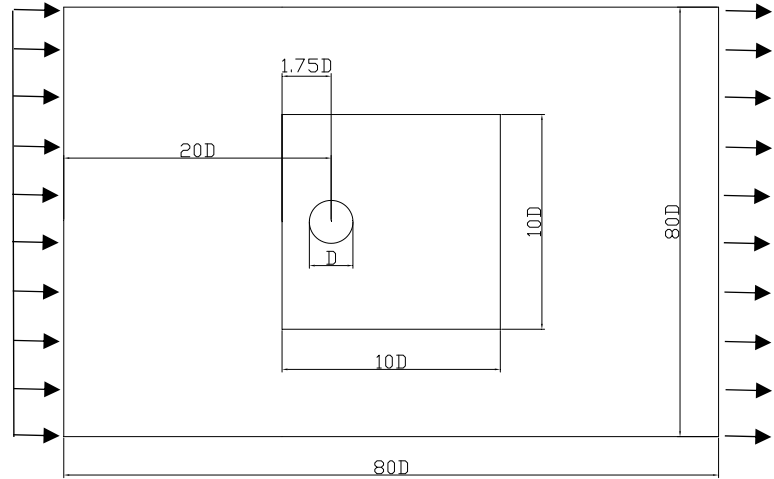

Figure 2. The schematic diagram of the computational domain for numerical simulations (whole domain) and the training domain for collecting datasets (area inside the dotted line).

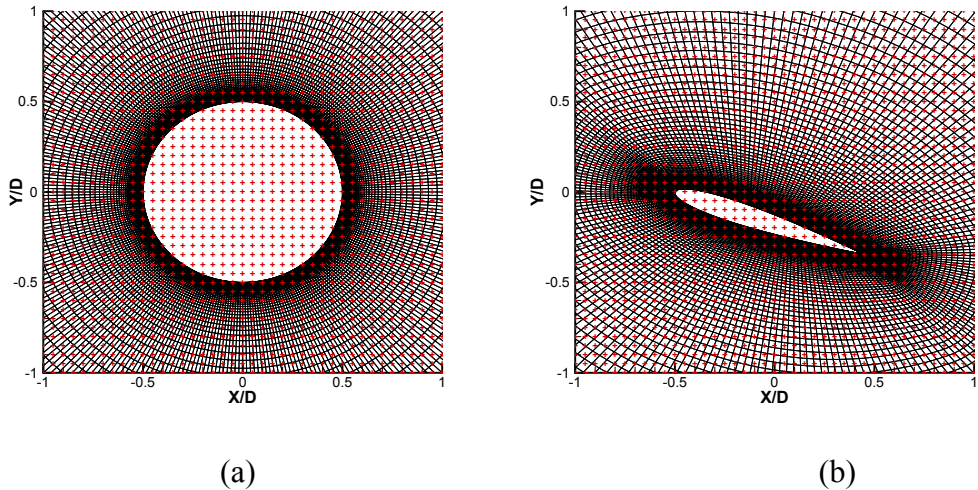

(a)             (b)

Figure 3. The discrete structured mesh representation of a cylinder and an airfoil (in black) on a Cartesian grid (in red).

Take the flow past a cylinder at Reynolds number Re = 200 experiment as example, detailing the construction process of the dataset. Flow are intended to be obtained by CFD calculation, and the computational area used is shown in Fig 2. The distance between the center of the body to the inlet is 20D (D is the diameter of the circular cylinder or the chord length of the airfoil) and to the outlet is 60D. The transverse width of the computational domain is 80D. A Dirichlet boundary condition is imposed on the inlet with the incoming velocity $U_0$; a free out-flow boundary condition is imposed on the outlet; a slip boundary condition is set up for the bottom and top boundaries of the flow; a no-slip boundary condition is set up for the solid body surface. By adjusting the solving time step length, make sure there are about 140 steps solution in each flow cycle. The simulation process captures the persistent behavior of the system. The original quantities in the incompressible Navier-Stokes equations are nondimensionalized as Eqn. 5. The dataset of each case consists of $N = 8000$ samples.

We use a rectangular area as the sampling area (area inside the dotted line) as shown in Fig 2. Lattice like sampling points of $200 \times 200$ are placed in the space, shown in Fig 3. Then, project the nondimensionalized flow field variables quantities onto the uniformly distributed grid. The values of the point inside the body are 0. Three-dimensional flow field variables ($p*$, $u*$ and $v*$) are extracted at each sampling point. The $200 \times 200 \times 3$ dimensional data that are extracted represent each instantaneous field. The obtained data is arranged in chronological order to obtain a dataset for training and testing the neural network.

Then, we train the proposed hybrid deep neural network in Tensorflow, with the values of the parameters for each layer as listed in Table 1. The learning rate used in optimization is set as 0.0005.

Table 1. Details of the structure parameters in the hybrid deep neural network.

| Layer | Kernel size/stride | Output size |
| --- | --- | --- |
| Conv1 | 2×2/1 | 16×200×200×4 |
| Conv2 | 3×3/2 | 16×100×100×8 |
| Conv3 | 3×3/2 | 16×50×50×16 |
| Conv4 | 3×3/2 | 16×25×25×32 |
| Conv5 | 3×3/2 | 16×13×13×64 |
| Conv6 | 3×3/2 | 16×7×7×64 |

| Reshape1 | ...... | 64×7×7×16 |
|---|---|---|
| ConvLSTM | 3×3/1 | 64×7×7×1 |
| Reshape2 | ...... | 1×7×7×64 |
| DeConv6 | 3×3/2 | 1×13×13×64 |
| DeConv5 | 3×3/2 | 1×25×25×32 |
| DeConv4 | 3×3/2 | 1×50×50×16 |
| DeConv3 | 3×3/2 | 1×100×100×8 |
| DeConv2 | 3×3/2 | 1×200×200×4 |
| DeConv1 | 2×2/1 | 1×200×200×3 |

Fig 4 shows that the training error defined by Eqn. 4 for three datasets decrease with the increasing number of training steps. After 700 training epochs, the training error converges to less than $5 \times 10^{-3}$.

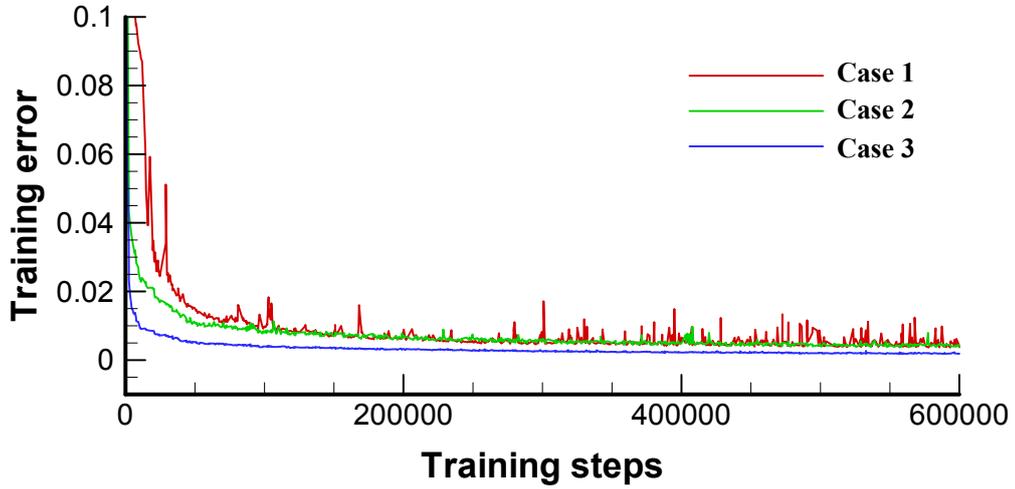

Figure 4. Training error for three datasets decrease with the increasing number of training steps.

Using the trained network, a flow field at time step *i* is recursively predicted based on the set of input flow field images at time step *i-1* to *i-k*. To achieve the goal of predicting the flow fields at future occasions based on flow fields at previous occasions continuously, we recycle the output of the trained network to its input and updated the input recursively as the time-step advancement, with an initial condition taken from $k$ snapshots of CFD data. So that, the network is able to achieve long-time predictions of flow fields even without known CFD data in the latter period.

## 4.2 The flow around a cylinder

We use the first two experiment to test the ability of the neural networks to predict unsteady flow in laminar and turbulent and reveal whether the neural network can capture spatio-temporal features of the unsteady flow at different states. The first numerical experiment (Case 1), the flow past a cylinder at Reynolds number Re = 200, is a laminar flow; and the second experiment (Case 2), the flow past a cylinder at Re = 4000, is a turbulent flow. They have different physical characteristics. Whether the same neural network structure can be applied to two different flow fields is the criterion for testing the applicability of the hybrid deep neural network.

Comparisons of instantaneous flow fields between the hybrid deep network predicted results and CFD results for two cases are shown in Fig. 5-8. Flow fields predicted and distributions of errors after a single time-step for two cases are shown in Fig. 5 and Fig. 7. Flow fields predicted and distributions of errors after 64 time-steps for two are shown in Fig. 6 and Fig. 8. From comparisons, we can get that flow fields predicted are found to agree well with CFD simulation flow fields for all cases.

Two characteristic positions, shown in Fig. 9, are selected to show the time series prediction accuracy. Time series of three flow field variables at these positions, predicted by the network and calculated by CFD, are compared in Fig. 10. All predicted results agree very well with those of CFD. It is proved again that each part of the neural network structure has completed the predetermined target and realized the prediction of the unsteady flow field. The neural network can not only predict the overall structural characteristics of the flow field, but also accurately predict the evolution process of the flow field. So we can conclude that the neural network can predict the spatio-temporal evolution of laminar and turbulent flow.

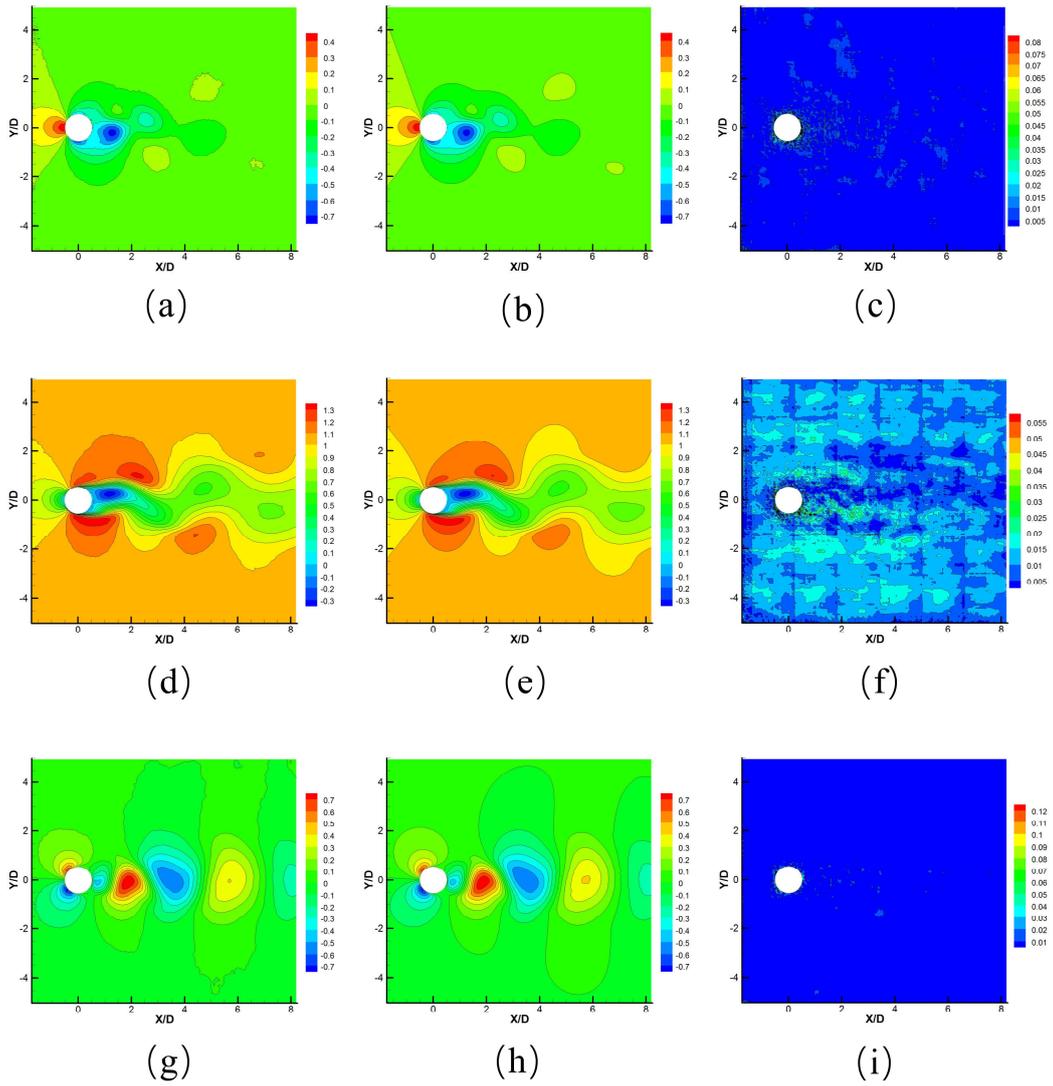

Figure 5. Comparisons of instantaneous flow fields after a single time-step between the model predictions and CFD results for Case 1 (flow past a cylinder at Re = 200): Network predictions for (a) pressure, (d) streamwise velocity, (g) vertical velocity; CFD results for (b) pressure, (e) streamwise velocity, (h) vertical velocity; Absolute prediction error for(c) pressure, (f) streamwise velocity, (i) vertical velocity.

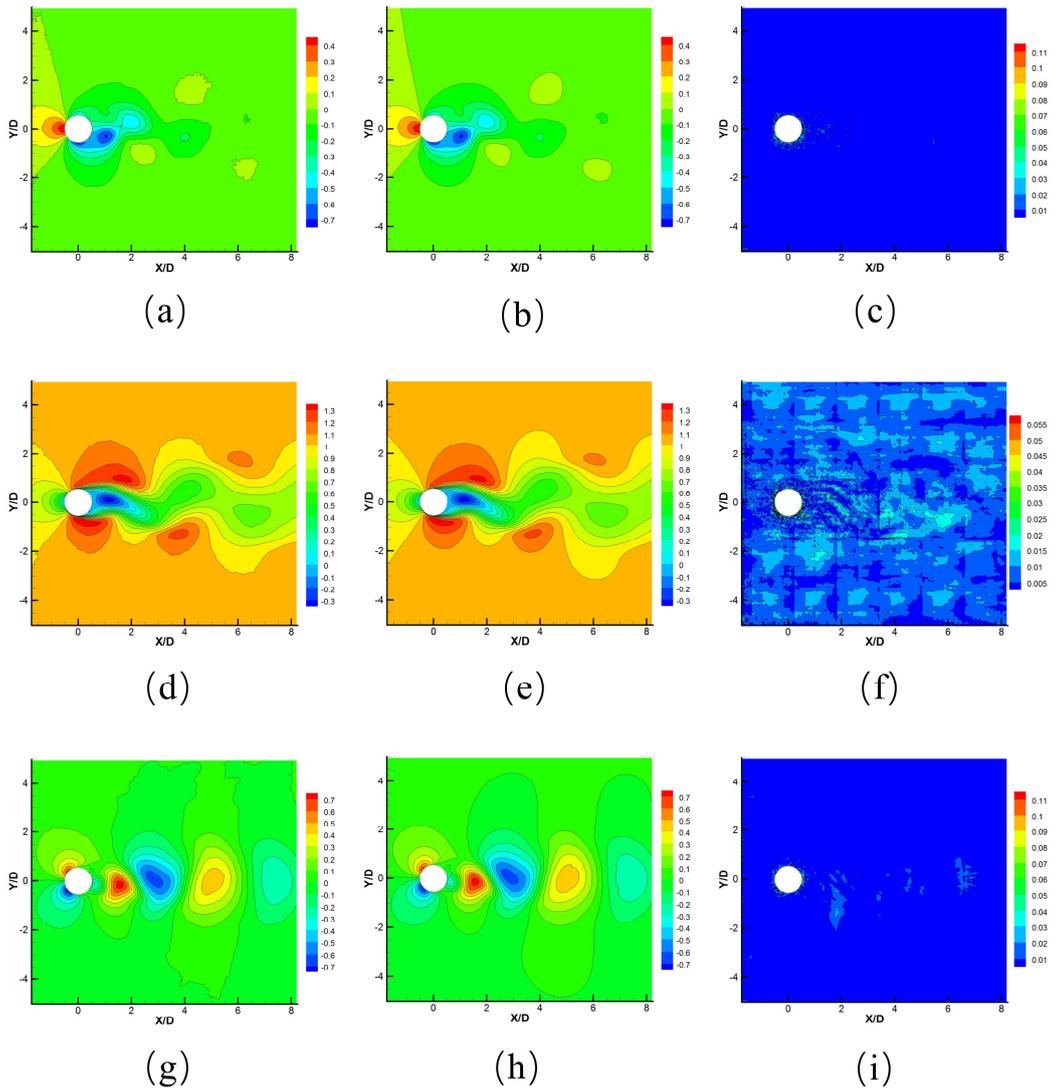

Figure 6. Comparisons of instantaneous flow fields after 64 time-steps between the model predictions and CFD results for Case 1 (flow past a cylinder at Re = 200): Network predictions for (a) pressure, (d) streamwise velocity, (g) vertical velocity; CFD results for (b) pressure, (e) streamwise velocity, (h) vertical velocity; Absolute prediction error for(c) pressure, (f) streamwise velocity, (i) vertical velocity.

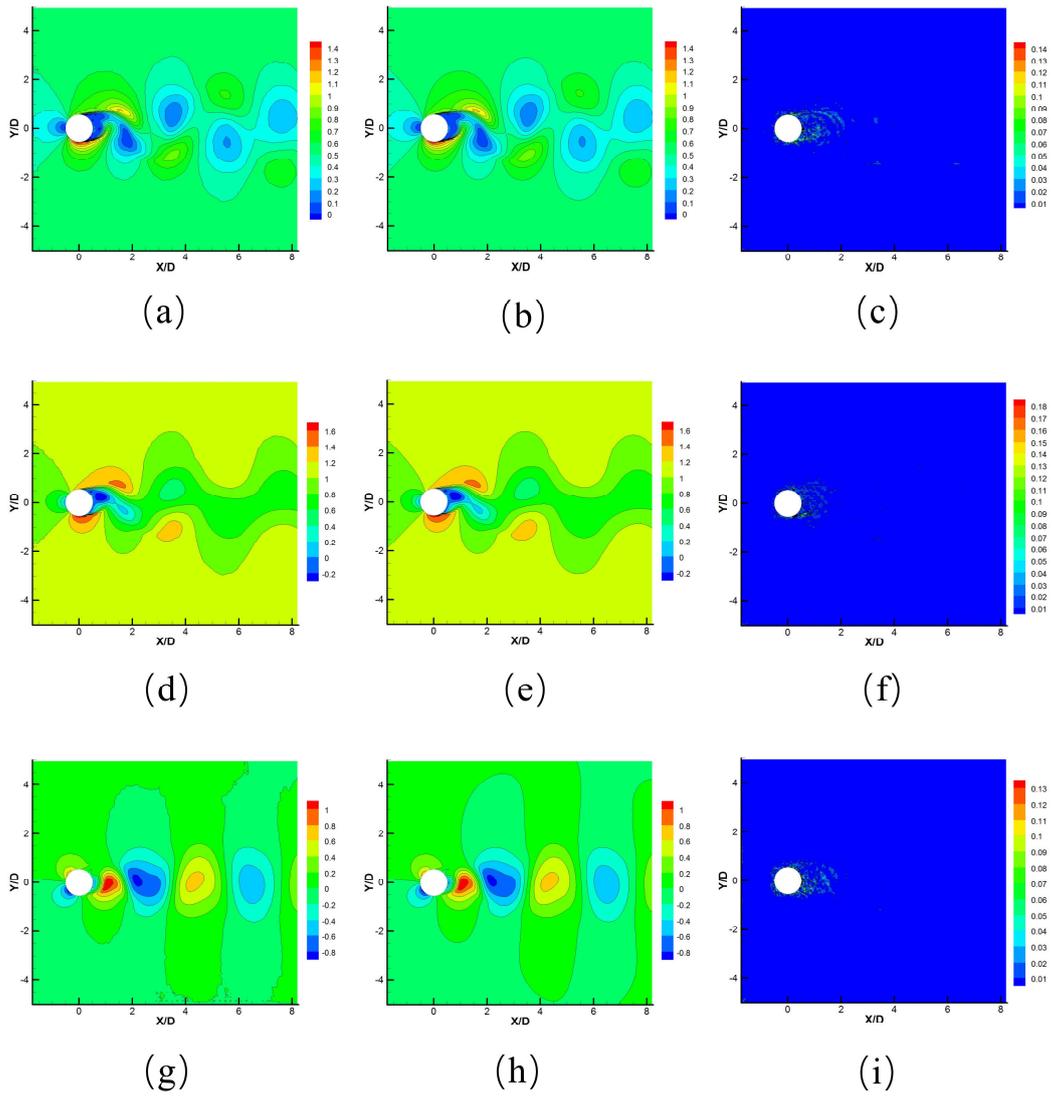

(a)  (b)  (c)

(d)  (e)  (f)

(g)  (h)  (i)

Figure 7. Comparisons of instantaneous flow fields after a single time-step between the model predictions and CFD results for Case 2 (flow past a cylinder at Re = 4000): Network predictions for (a) pressure, (d) streamwise velocity, (g) vertical velocity; CFD results for (b) pressure, (e) streamwise velocity, (h) vertical velocity; Absolute prediction error for(c) pressure, (f) streamwise velocity, (i) vertical velocity.

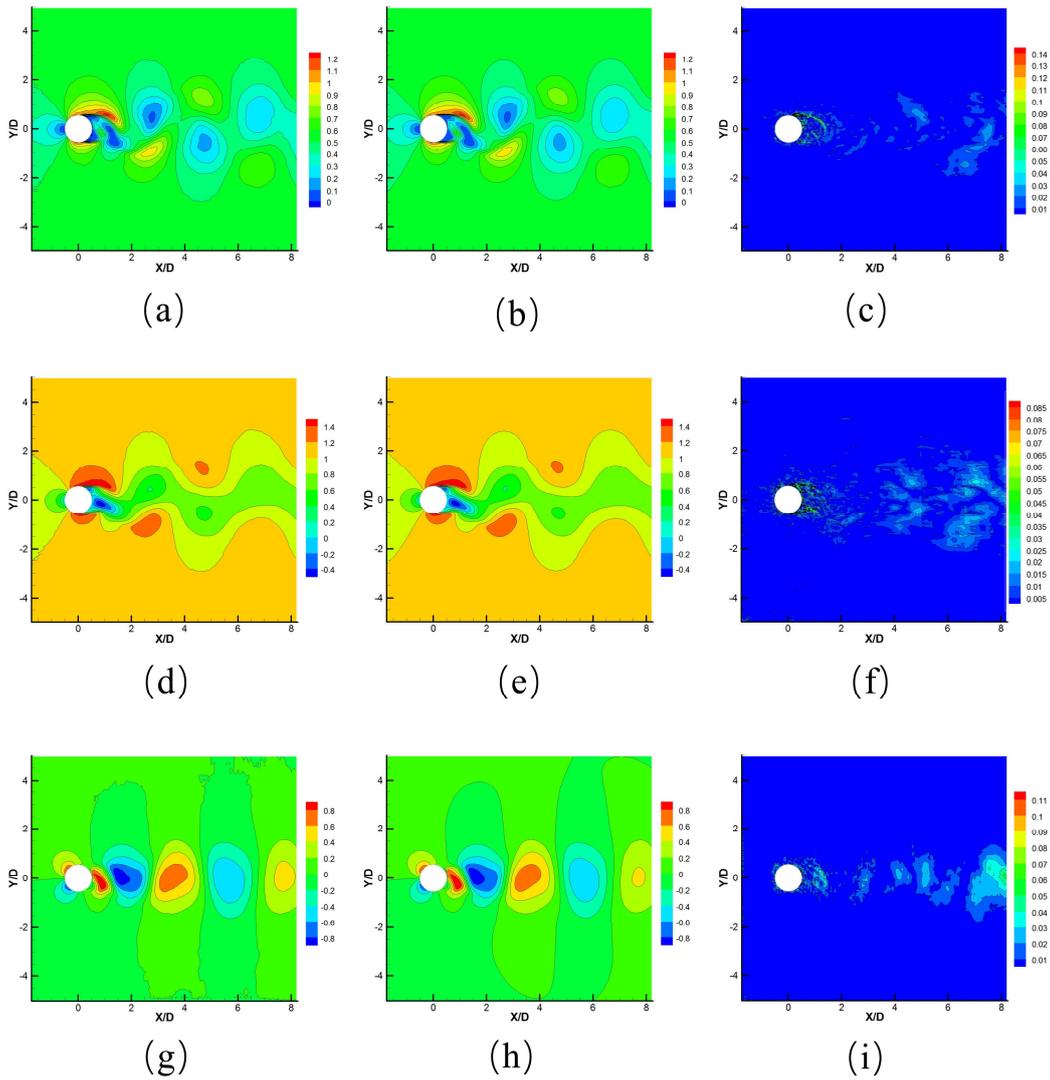

(a)          (b)          (c)

(d)          (e)          (f)

(g)          (h)          (i)

Figure 8. Comparisons of instantaneous flow fields after 64 time-steps between the model predictions and CFD results for Case 2 (flow past a cylinder at Re = 4000): Network predictions for (a) pressure, (d) streamwise velocity, (g) vertical velocity; CFD results for (b) pressure, (e) streamwise velocity, (h) vertical velocity; Absolute prediction error for (c) pressure, (f) streamwise velocity, (i) vertical velocity.

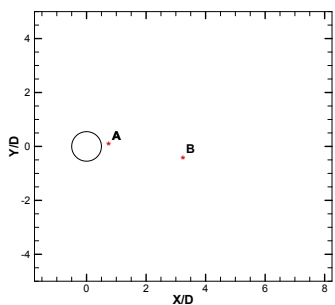

Figure. 9. A map of the positions selected in the wake to show the model prediction accuracy. The spatial coordinates of these points are described by dimensionless x* and y*: circle center (0, 0); point in the separating free layer A (0.75, 0), point in the wake C (3.25, -0.5).

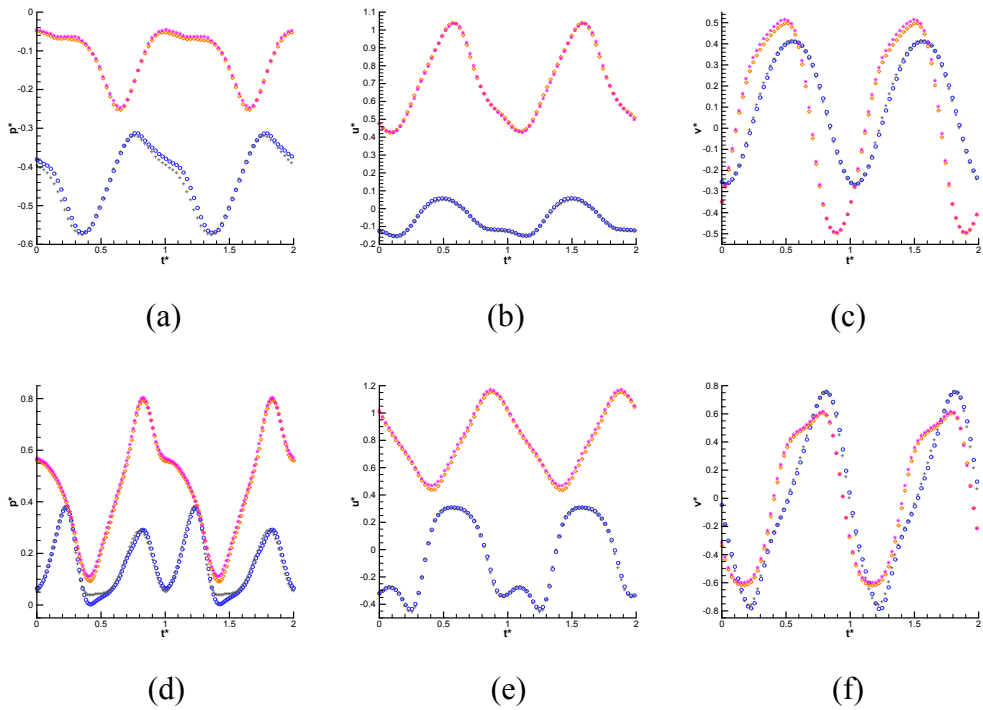

(a)             (b)             (c)

(d)             (e)             (f)

Figure 10. Comparisons of velocity time histories in the wake flow between the model predictions and CFD results for all test cases (circle shape point, model prediction, $x^* = 0.75$, $y^* = 0$; plus shape point, CFD results, $x^* = 0.75$, $y^* = 0$; diamond shape point, model prediction, $x^* = 3.25$, $y^* = -0.5$; star shape point, CFD results, $x^* = 3.25$, $y^* = -0.5$. Case 1 for (a) pressure, (b) streamwise velocity, (c) vertical velocity; Case 2 for (d) pressure, (e) streamwise velocity, (f) vertical velocity.

## 4.3 The flow around an airfoil

In order to further explore the applicability of the neural network structure, the neural network is used to predict the more complex unsteady flow field, the flow around an airfoil at higher Reynolds number. Comparisons of instantaneous flow fields between the hybrid deep network predicted results and CFD results are shown in Fig. 11 and Fig. 12. Flow fields predicted are found to agree well with CFD simulation flow fields. Time series of three flow field variables at two selected positions, predicted by the model and calculated by CFD, respectively, are compared in Fig. 13. All predicted results agree very well with those of CFD. And the error accumulation on differences between the predicted and the CFD simulation flow fields as the number of the recursive step increases is not obvious. The hybrid deep neural network can predict the spatial and temporal evolution of turbulence flow over a complex shape (airfoil). We treat the input-output as images, of which the different characteristics are easy to be captured by CNN layer. So the hybrid deep neural network is suitable for learning the spatio-temporal features of different unsteady flow.

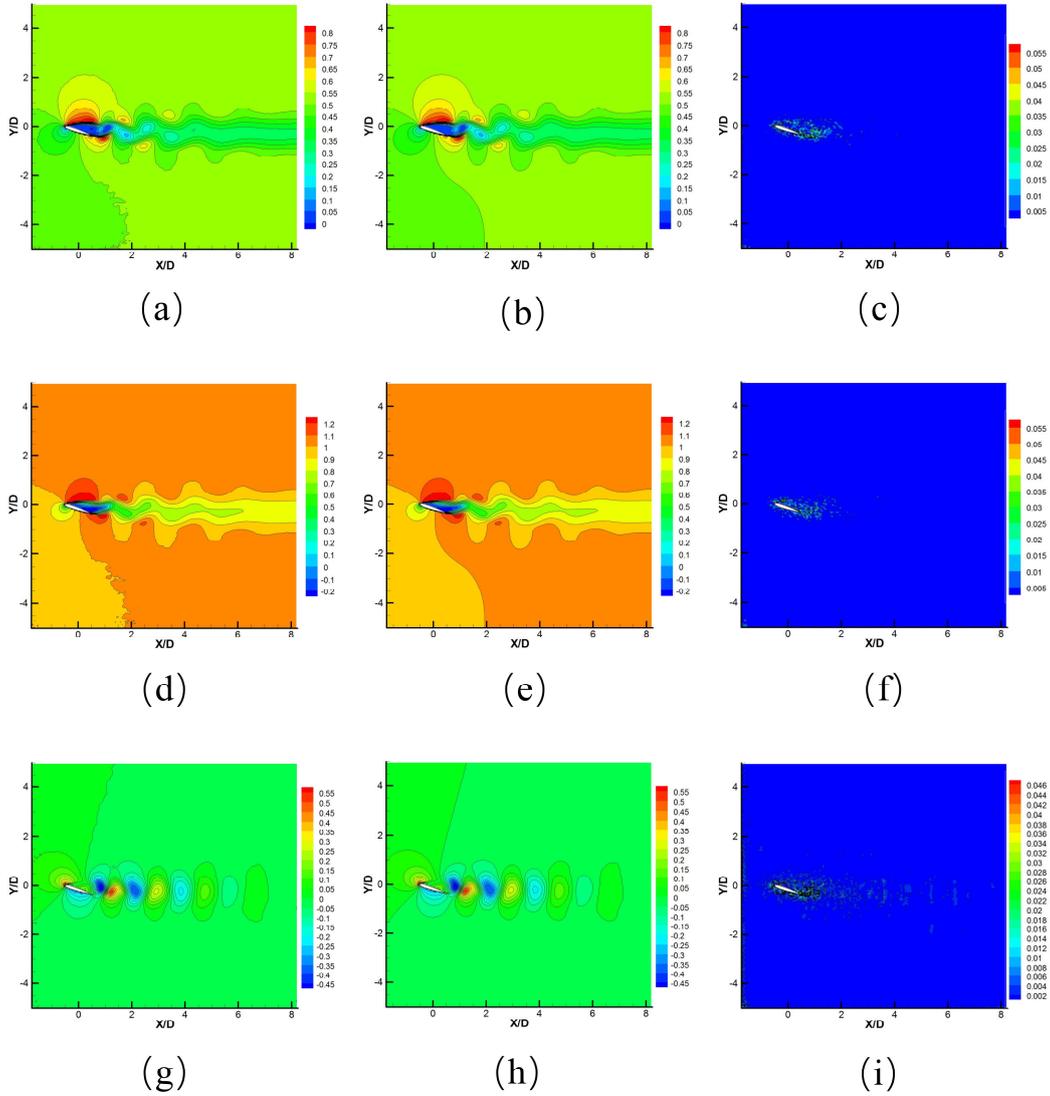

Figure 11. Comparisons of instantaneous flow fields after a single time-step between the model predictions and CFD results for Case 3 (flow past an airfoil at Re = 8000): Network predictions for (a) pressure, (d) streamwise velocity, (g) vertical velocity; CFD results for (b) pressure, (e) streamwise velocity, (h) vertical velocity; Absolute prediction error for(c) pressure, (f) streamwise velocity, (i) vertical velocity.

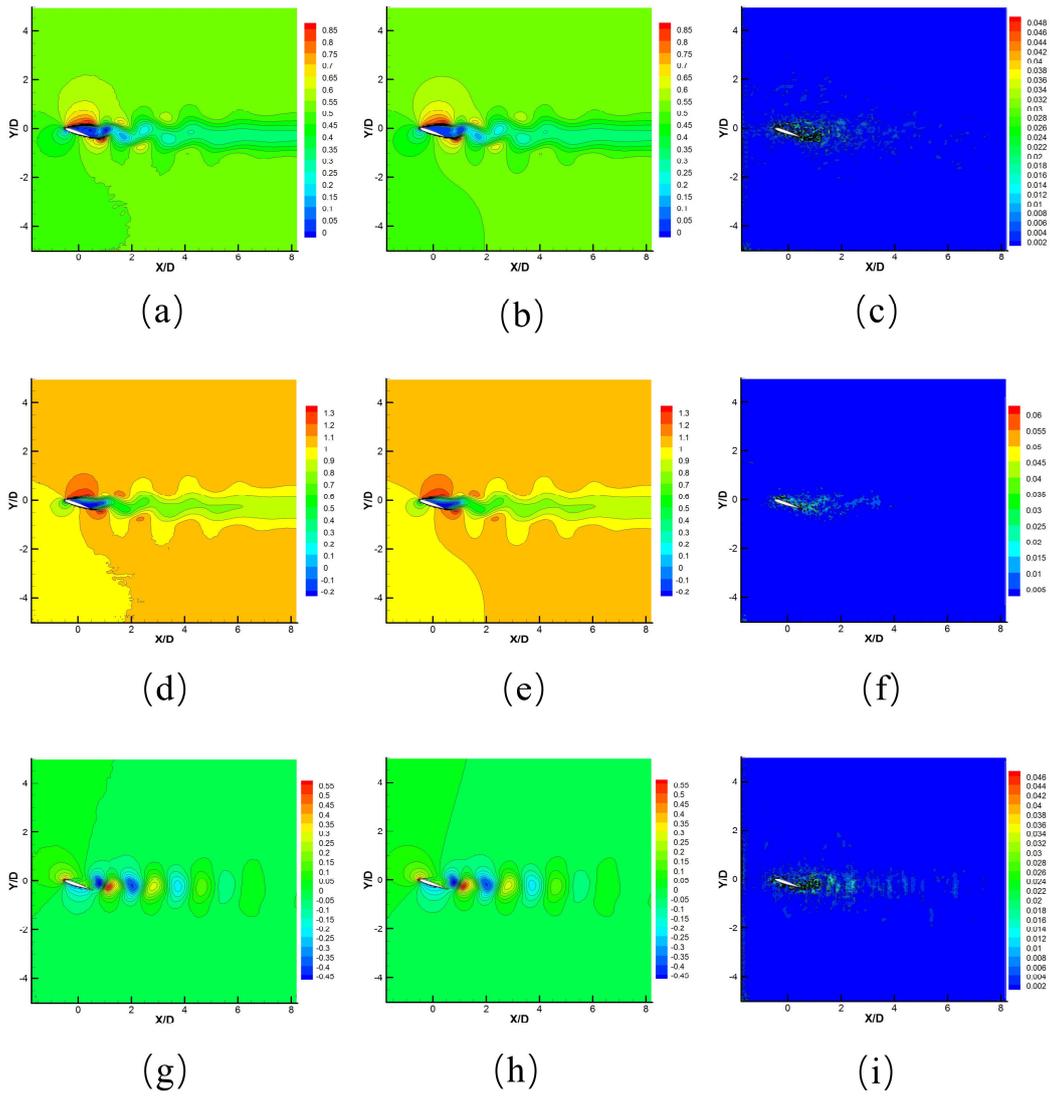

Figure 12. Comparisons of instantaneous flow fields after 64 time-steps between the model predictions and CFD results for Case 3 (flow past an airfoil at Re = 8000): Network predictions for (a) pressure, (d) streamwise velocity, (g) vertical velocity; CFD results for (b) pressure, (e) streamwise velocity, (h) vertical velocity; Absolute prediction error for(c) pressure, (f) streamwise velocity, (i) vertical velocity.

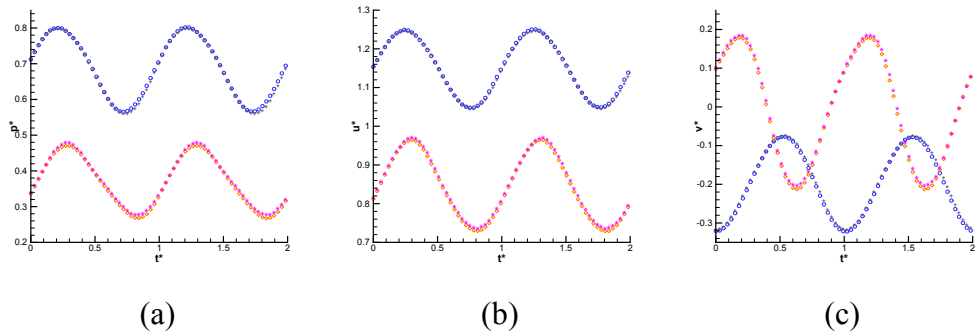

|          |          |          |
|:--------:|:--------:|:--------:|
|   (a)    |   (b)    |   (c)    |

Figure 13. Comparisons of velocity time histories in the wake flow between the model predictions and CFD results for all test cases (circle shape point, model prediction, x* = 0.75, y* = 0; plus shape point, CFD results, x* = 0.75, y* = 0; diamond shape point, model prediction, x* = 3.25, y*= -0.5; star shape point, CFD results, x* = 3.25, y*= -0.5. Case 3 for (a) pressure, (b) streamwise velocity, (c) vertical velocity.

## 5. Conclusions

This work tried to use one neural network to reduce the dimensionality of the unsteady flow fields and capture the spatio-temporal features of flow field series simultaneously. A novel hybrid deep neural network architecture was designed to capture the unsteady flow spatio-temporal features directly from the high-dimensional unsteady flow fields. The flow around a cylinder at various Reynolds numbers and the flow around an airfoil at higher Reynolds number are carried out to establish the datasets training the networks separately. The trained hybrid DNNs were then tested by the prediction of the flow fields at future occasions. The predicted flow fields using the trained hybrid deep neural networks are in good agreement with the flow fields calculated directly by the computational fluid dynamic solver. The hybrid deep neural network can achieve fast and accurate prediction of unsteady flow fields, which is very important for flow control. The following conclusions are obtained from this study.

The hybrid deep neural network is constituted by CNN, ConvLSTM and DeCNN. The CNN layers are designed to capture the complex mapping directly from the high-dimensional input field and represent it in low-dimensional form. The ConvLSTM layer is designed to capture temporal features between low-dimensional features representation and predict the low-dimensional features of flow fields at future occasions. The DeCNN layers are used to represent the predicted low-dimensional features to high-dimensional output field, with the same dimension as input field. This kind of neural network can capture accurate spatial-temporal information from the spatial-temporal series of unsteady flow. Numerical simulations of flow around a cylinder at various Reynolds numbers and flow around an airfoil are performed to obtain the training and testing dataset. Instantaneous flow fields and velocity time histories predicted are found to agree well with CFD simulation flow fields for all cases. The results show, the difference between different flows caused by different object shape and Reynolds number have little effect on model prediction accuracy, since we treat the input-output as images. The different characteristics of different images is easy to be captured by CNN layer. All test cases show that the maximum errors are located accelerating boundary layers on the cylinder wall or in the wake, where the flow field changes drastically, but the maximum errors are in a low level. The prediction accuracy of the hybrid deep neural networks is acceptable. The new prediction method can be used in fluid-structure interactions and flow control, where the fast high-dimensional nonlinear unsteady flow calculation is needed.

This study show the potential capability of the novel hybrid deep neural network based reduced order model in the fast prediction of the unsteady flow. In future work, we will improve the network structure so that it can learn unsteady flow data in different Reynolds numbers and body shapes at the same time. The ultimate goal is to predict flow fields at the Reynolds numbers and body shapes, which are not included in training dataset. And the new method of using the hybrid deep neural network in fluid-structure interactions and flow control is also worthy of further study.

## Acknowledgments

This work was partially supported by the National Natural Science Foundation of China (No. 11672225, 11872293), and the Key Laboratory of Aerodynamics Noise Control (No.1801ANCL20180103).

# Reference

1   F.Fang, C.Pain, I.M. Navon, A.H. Elsheikh, J. Du, and D.Xiao. Non-linear Petrov-Galerkin methods for Reduced Order Hyperbolic Equations and Discontinuous Finite Element Methods. Journal of Computational Physics, 234:540–559 (2013_.

2   J. L. Lumley, The structure of inhomogeneous turbulent flows, Atmospheric Turbulence and Radio Wave Propagation 790, 166–178 (1967).

3   C. W. Rowley, I. Mezic, S. Bagheri, P. Schlatter, and D. S. Henningson, Spectral analysis of nonlinear flows, Journal of Fluid Mechanics 641, 115–127 (2009).

4   Yann LeCun, Yoshua Bengio, and Geoffrey Hinton. Deep learning. Nature, 521(7553):436–444 (2015).

5   Tara N Sainath, Abdel-rahman Mohamed, Brian Kingsbury, and Bhuvana Ramabhadran. Deep convolutional neural networks for lvcsr. In Acoustics, speech and signal processing (ICASSP), 2013 IEEE international conference on, 8614–8618. IEEE (2013).

6   Alex Krizhevsky, Ilya Sutskever, and Geoffrey E Hinton. Imagenet classification with deep convolutional neural networks. In Advances in neural information processing systems, pages 1097–1105 (2012).

7   Hui Y Xiong, Babak Alipanahi, Leo J Lee, Hannes Bretschneider, Daniele Merico, Ryan KC Yuen, Yimin Hua, Serge Gueroussov, Hamed S Najafabadi, Timothy R Hughes, et al. The human splicing code reveals new insights into the genetic determinants of disease. Science, 347(6218):1254806 (2015).

8   J. Ling, A. Kurzawski, and J. Templeton. Reynolds averaged turbulence modelling using deep neural networks with embedded invariance. Journal of Fluid Mechanics, vol. 807, pp. 155–166, (2016).

9   Wu J, Xiao H, Paterson E. Physics-informaed machine learning approach for augmenting turbulence models: A comprehensive framework. Physical Review Fluids, 3, 074602 (2018).

10  R. Maulik, O. San, A. Rasheed, and P. Vedula. Subgrid modelling for two-dimensional turbulence using neural networks. Journal of Fluid Mechanics, vol.

858, pp. 122–144 (2019).

11  Mingliang Wang, Han-Xiong Li, Xin Chen, and Yun Chen. Deep learning-based model reduction for distributed parameter systems. IEEE Transactions on Systems, Man, and Cybernetics: Systems, 46(12):1664–1674 (2016).

12  Noriyasu Omataa, Susumu Shirayama. A novel method of low-dimensional representation for temporal behavior of flow fields using deep autoencoder. AIP Advances 9, 015006 (2019).

13  Kai Fukami1, Koji Fukagata, Kunihiko Taira. Super-resolution reconstruction of turbulent flows with machine learning. arXiv preprint arXiv: 1811.11328 (2019).

14  Xiaowei Jin, Peng Cheng, Wen-Li Chen, Hui Li. Prediction model of velocity field around circular cylinder over various Reynolds numbers by fusion convolutional neural networks based on pressure on the cylinder. Phys. Fluids 30, 047105 (2018).

15  Arvind T. Mohan, Datta V. Gaitonde. A Deep Learning based Approach to Reduced Order Modeling for Turbulent Flow Control using LSTM Neural Networks. arXiv preprint arXiv:1804.09269 (2018).

16  Arvind T. Mohan, Don Daniel, Michael Chertkov, Daniel Livescu. Compressed Convolutional LSTM: An Efficient Deep Learning Framework to Model High Fidelity 3D Turbulence. arXiv preprint arXiv:1903.00033 (2019).

17  Sangseung Lee, Donghyun You. Data-driven prediction of unsteady flow over a circular cylinder using deep learning. arXiv preprint arXiv:1804.06076v1 (2018).

18  Mario Ruttgers, Sangseung Lee, and Donghyun You. Prediction of typhoon tracks using a generative adversarial network with observational and meteorological data. arXiv preprint arXiv:1812.01943 v1 (2018).

19  Xingjian Shi Zhourong Chen Hao Wang Dit-Yan Yeung. Convolutional LSTM network: A machine learning approach for precipitation nowcasting. In Advances in neural information processing systems, pp. 802–810 (2015).

20  Zeiler, Matthew D., et al. Deconvolutional networks. 2010 IEEE Computer Society Conference on Computer Vision and Pattern Recognition IEEE (2010).

21  Zeiler, Matthew D., G. W. Taylor, and R. Fergus. Adaptive deconvolutional networks for mid and high level feature learning. 2011 International Conference on Computer Vision IEEE (2012).

22   J. Bouvrie. Notes on convolutional neural networks. Massachusetts Institute of Technology (2006).

23   D. P. Kingma and J. Ba. Adam: A method for stochastic optimization. arXiv preprint arXiv:1412.6980 (2015).

24   M. Abadi, A. Agarwal, P. Barham, E. Brevdo, Z. Chen, C. Citro, G. S.Corrado, A. Davis, J. Dean,M. Devin, S. Ghemawat, I. Goodfellow, A. Harp, G. Irving,M. Isard, R. Jozefowicz, Y. Jia,M. Kudlur, L. Kaiser, J. Levenberg, D. Mane, M. Schuster, S. Moore, R. Monga, D. Murray, C. Olah, J. Shlens, I. Sutskever, B. Steiner, K. Talwar, P. Tucker, V. Vasudevan, V. Vanhoucke, F. Viegas, P. Warden, O. Vinyals, M. Wattenberg, M. Wicke, X. Zheng, and Y. Yu, "TensorFlow: Large-scale machine learning on heterogeneous systems," software available from http://tensorflow.org/ (2015).